\documentclass[12pt]{article}
\usepackage{epsfig}

\newcommand{\be}{\begin{equation}}
\newcommand{\ee}{\end{equation}}
\newcommand{\beqn}{\begin{eqnarray}}
\newcommand{\eeqn}{\end{eqnarray}}
\newcommand{\eq}[1]{(\ref{#1})}

\newcommand{\itep}
{~\vspace{-1.5cm}

\begin{flushright}
{\large LU-ITP 2003/001}\\
{\large KANAZAWA-03-03}\\
{\large ITEP-LAT/2003-01}\\
{\large HU-EP-02-61}\\
{\large January 10, 2003}
\end{flushright}
\vspace{1.0cm}}

\begin{document}

\thispagestyle{empty}

\baselineskip=14pt

\begin{center}

\itep

{\Large\bf
Confinement, deconfinement and the photon propagator
in 3D \lowercase{c}QED on the lattice
}

\vskip 1.0cm
{\large
M.~N.~Chernodub$^{a,b,}$\footnote{\uppercase{W}ork
supported by \uppercase{JSPS} \uppercase{F}ellowship
\uppercase{P}01023.},
E.--M.~Ilgenfritz$^c$ and A.~Schiller$^{d,}$\footnote{
Talk presented by A.S. at Confinement 5, Gargnano, September 2002
}
}\\

\vspace{.4cm}

{ \it
$^a$ ITEP, B. Cheremushkinskaya 25, Moscow, 117259, Russia

\vspace{0.3cm}

$^b$ Institute for Theoretical Physics, Kanazawa University,\\
Kanazawa 920-1192, Japan

\vspace{0.3cm}

$^c$ Institut f\"ur Physik, Humboldt--Universit\"at zu Berlin,
D-10115 Berlin, Germany

\vspace{0.3cm}
$^d$ Institut f\"ur Theoretische Physik and NTZ, Universit\"at
Leipzig,\\ D-04109 Leipzig, Germany
}
\end{center}

\begin{abstract}
{We report on a lattice study of the
gauge boson propagator of $3D$ compact QED
in Landau gauge at zero and finite temperature.
Non-perturbative effects are reflected by the generation
of a mass $m$ and by an anomalous dimension $\alpha$.
These effects can be attributed
to monopoles and are absent in the propagator of the
regular part of the gauge field.
}
\end{abstract}

Three--dimensional compact electrodynamics (cQED$_3$) shares
two features with QCD, confinement and chiral symmetry breaking.
Confinement of electrically charged particles is caused by a
plasma of monopoles~\cite{Polyakov}.
In this contribution we report how confinement
is manifest in the gauge boson propagator of $3D$ cQED
at zero and finite temperature\cite{CISLetter,Chernodub:2002gp}.

For our lattice study we have adopted the Wilson action,
$S[\theta] = \beta \sum_p \left( 1 - \cos \theta_p \right)$, where
$\theta_p$ is the
field strength corresponding to the
$U(1)$ link field $\theta_l$. The coupling $\beta$ is
related to the lattice spacing $a$ and the coupling
$g_3$ of the $3D$ continuum theory, $\beta = 1 \slash (a\, g^2_3)$.
The photon propagator is measured in Landau gauge.
This gauge is defined as a global maximum of the
functional $\sum_l \cos\theta^{G}_l$ with
respect to gauge transformations $G$.
To select the best gauge realization, we have
evaluated $N_G$ local maxima of that
functional and studied the dependence of the propagator on $N_G$.
Simulations for $T=0$ have been performed on a $32^3$ lattice,
those for $T>0$ on a $32^2 \times 8$ lattice.
For the Monte Carlo algorithm
and the gauge fixing procedure see
Ref. \cite{Chernodub:2002gp}

The gauge field in lattice position space
is taken as
$A_{{\vec n}+\frac{1}{2}{\vec \mu},\mu}
= \sin \theta_{{\vec n},\mu} /(g_3\,a)$.
Then we define the gauge boson propagator in lattice
momentum space
by $D_{\mu\nu}({\vec p})
= \langle \tilde{A}_{ {\vec k},\mu} \tilde{A}_{-{\vec k},\nu} \rangle$
in terms of the Fourier transformed gauge potential
for $p_{\nu} = (2/a) \sin (2 \pi k_{\nu}/L_{\nu})$.
The most general tensor structure of $D_{\mu\nu}$ at $T=0$ is given by
$$
  D_{\mu\nu}(\vec p)= P_{\mu\nu}(\vec p) D (p^2)
  +  p_\mu p_\nu  \,  F(p^2)/(p^2)^2
$$
with the $3D$ transverse projection operator
$P_{\mu\nu}(\vec p)= \delta_{\mu\nu}- (p_\mu p_\nu)/{p^2}$.
The transverse and longitudinal propagators, $D$ and $F$, are extracted by
projection and are found approximately rotationally invariant or vanishing.
For $T>0$
we have to consider two scalar functions
$D_{T/L}$ using $2D$ transverse/longitudinal
projection operators $P_{\mu\nu}^{T/L}$
(in $2+1$ dimensions both are transverse):
$$
  D_{\mu\nu}(\vec p)= P^T_{\mu\nu} D_T + P^L_{\mu\nu} D_L
  + p_\mu p_\nu  \, F/(p^2)^2 \,.
$$

To discuss the propagators from the monopole plasma
point of view, we
have decomposed the gauge fields links into singular
(monopole) and regular (photon) contributions\cite{PhMon,Chernodub:2002gp}
$\theta_l = \theta_l^{phot} + \theta_l^{mono}$.
Once that decomposition is achieved, one is
in the position to study monopole,
photon and mixed contributions of $D$ or $D_{L/T}$ (and $F$), separately.

\begin{figure}[!htb]
\vspace{7mm}
  \begin{center}
    \begin{tabular}{cc}
      \epsfxsize=7.5cm \epsffile{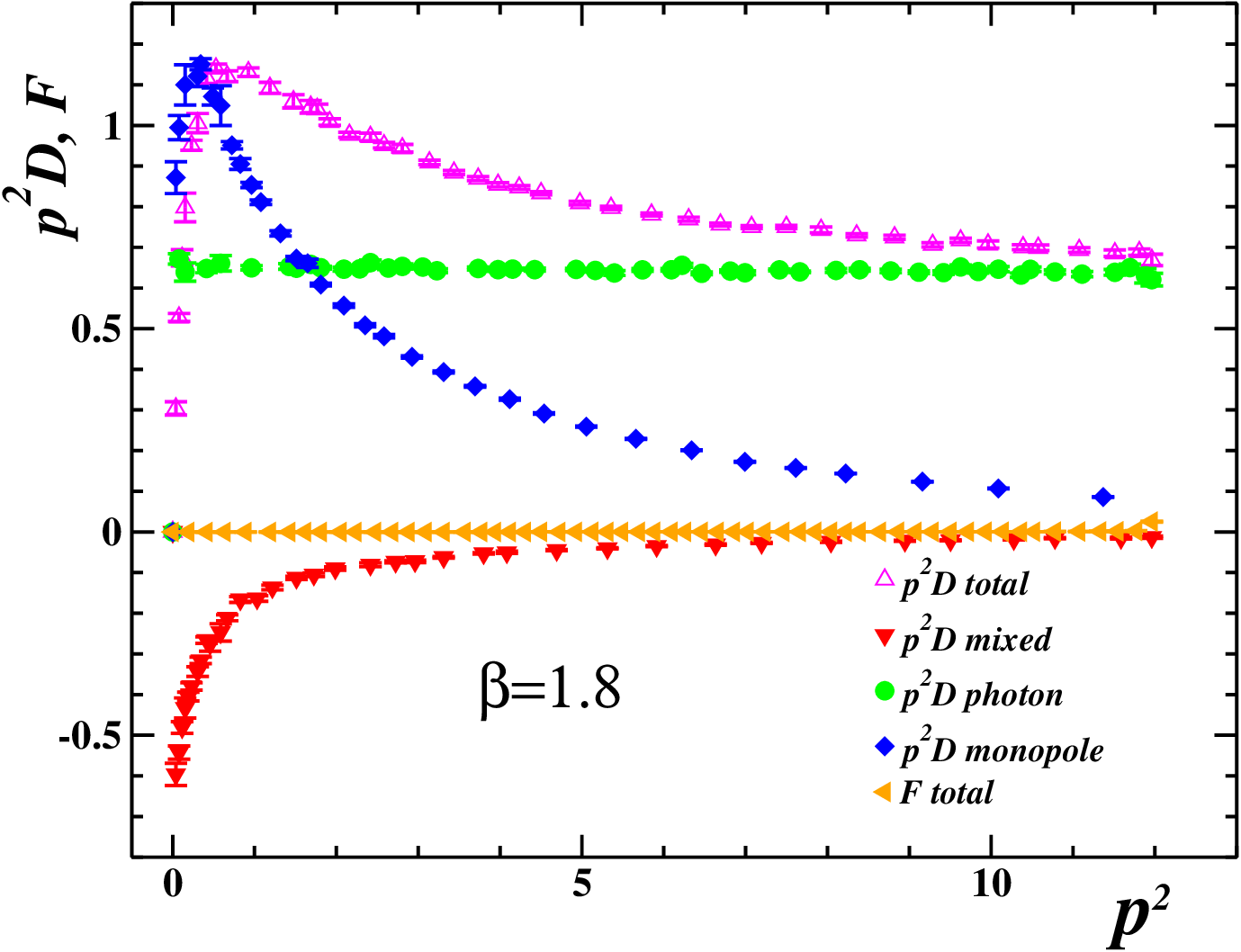} &
      \hspace{5mm} \epsfxsize=7.1cm \epsffile{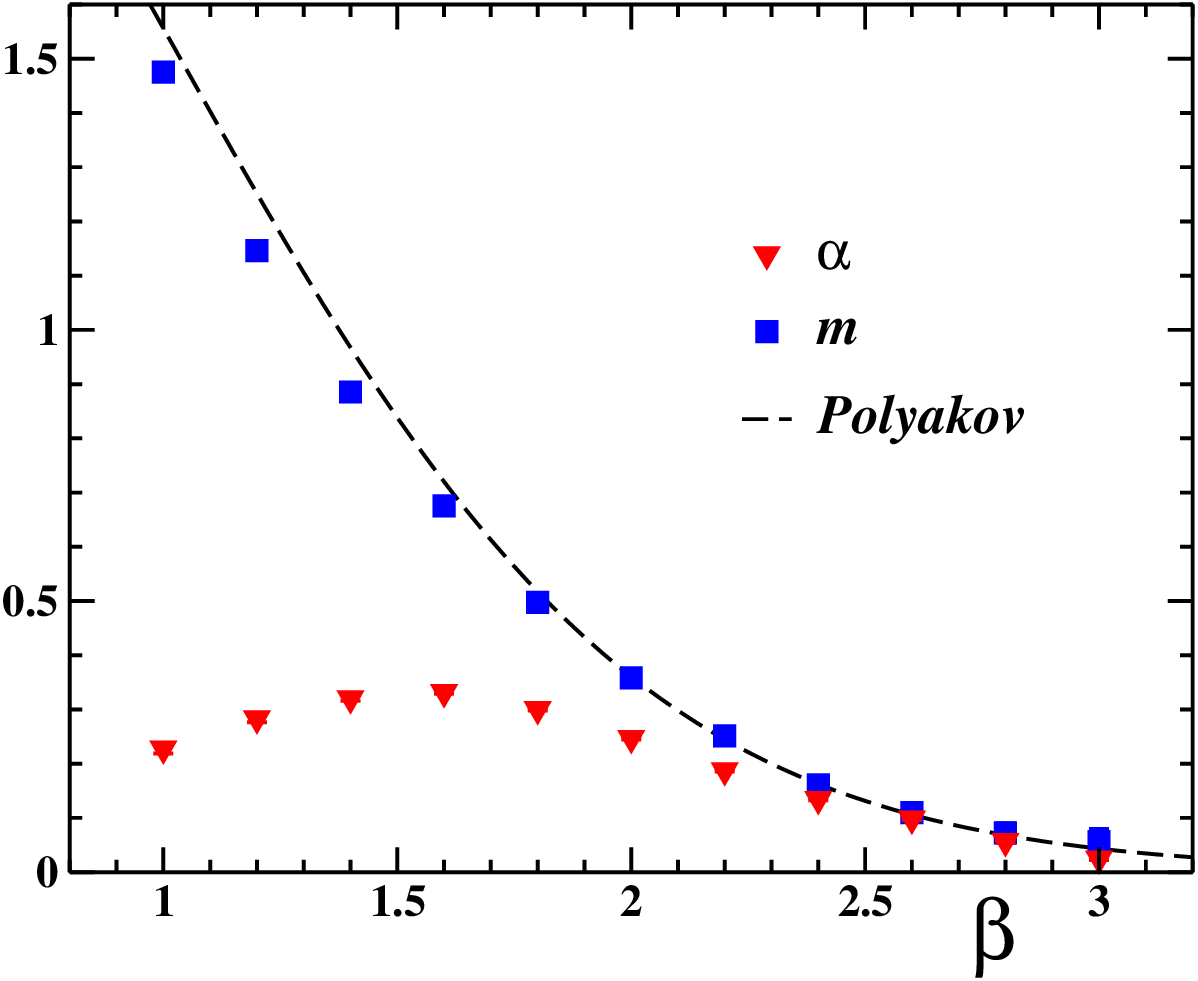} \\
      (a) & \hspace{5mm}  (b)
\vspace{-3mm}
    \end{tabular}
  \end{center}
  \caption{(a) The propagator $p^2 D$ and its contributions, as well as
             the (vanishing) propagator $F$ vs. $p^2$ at $\beta=1.8$.
             (b) Fitted $\alpha$ and $m$ vs. $\beta$ for $D$.}
  \label{fig1}
\end{figure}

As an example, we show in Fig.~\ref{fig1}~(a)
at $T=0$ the transverse and longitudinal propagators (multiplied by $p^2$)
together with
the decomposition of $D$.
The monopole part $D^{mono}$ reaches a maximum in
the low momentum region before it drops towards $p^2=0$.
In contrast to the total $D$, the regular (photon) part is singular at
$p^2=0$, and $p^2 D^{phot}$ is flat vs. $p^2$.

Following Ref.\cite{CISLetter} we describe $D$ by the function
\beqn
  D(p^2) = ({Z}/{\beta}) \,
 {m^{2\alpha}}/[{p^{2(1+\alpha)}+m^{2(1+\alpha)}}]
  + C \,.
  \label{def:anomalous_fit}
\eeqn
The photon part $D^{phot}$ is fitted using (\ref{def:anomalous_fit})
with $\alpha = m = 0$.
At finite $T$, the propagator data for $D_L$ and $D_T$
are analyzed for $p_3=0$, as a function of
${\mathbf p}^2$ using the same fit function~\eq{def:anomalous_fit}.

\begin{figure}[!htb]
\vspace{7mm}
  \begin{center}
    \begin{tabular}{cc}
      \epsfxsize=7.5cm \epsffile{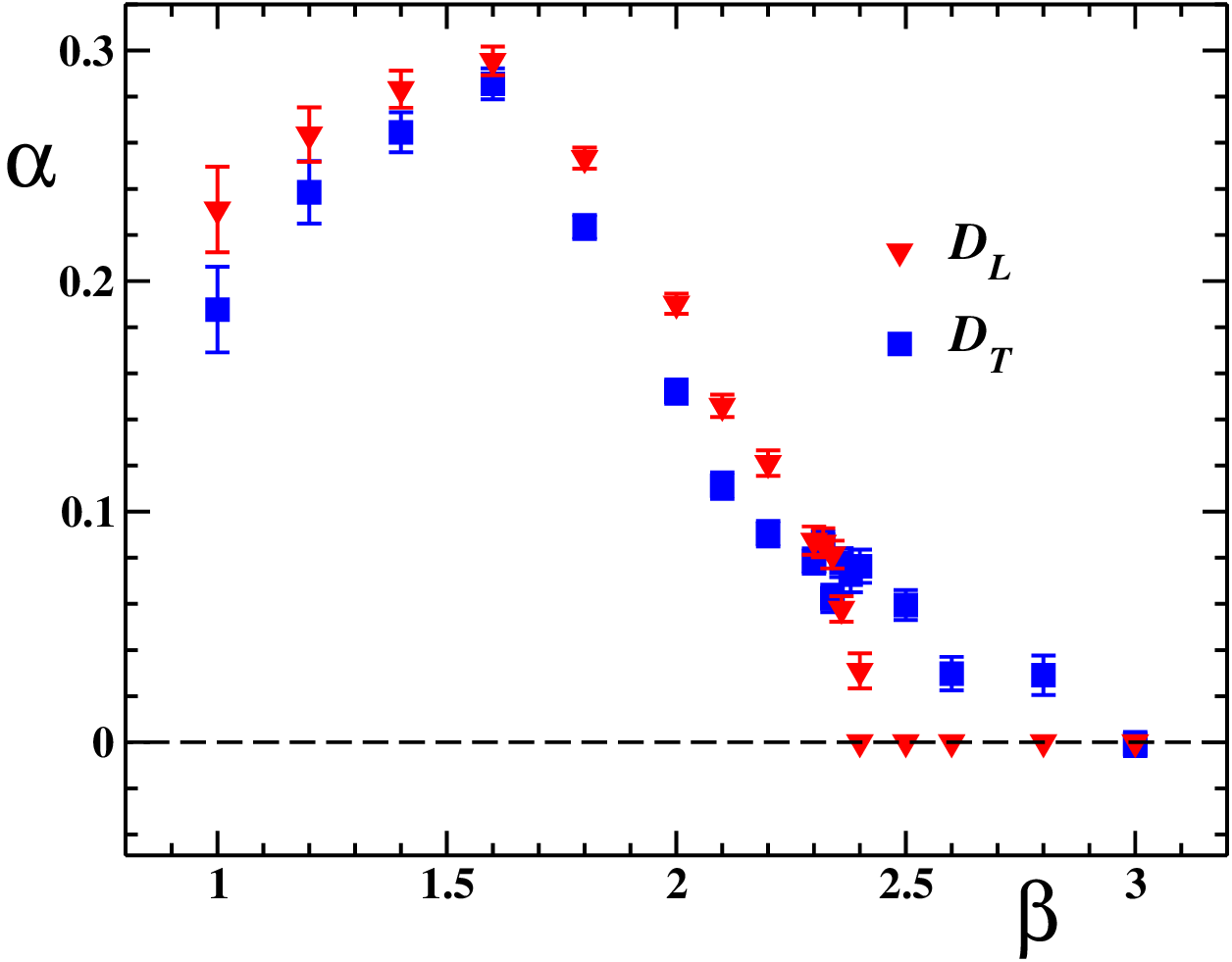} &
      \hspace{5mm} \epsfxsize=7.5cm \epsffile{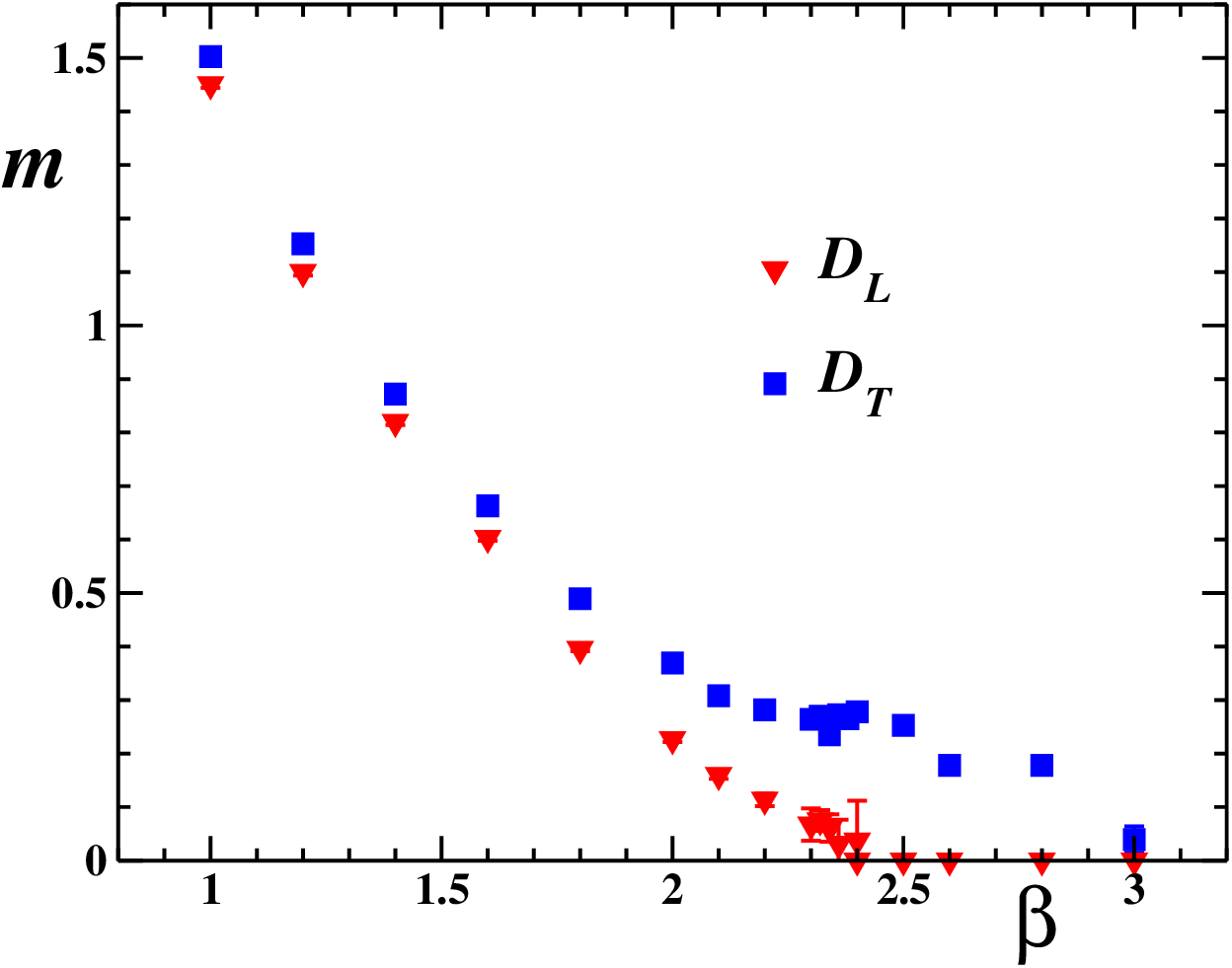} \\
      (a) & \hspace{5mm}  (b)
\vspace{-3mm}
    \end{tabular}
  \end{center}
  \caption{Fitted $\alpha$ and $m$ vs. $\beta$ for
           $D_L$ and $D_T$.}
  \label{fig2}
\end{figure}

The fit parameters $\alpha$ and $m$ of the
propagators $D$, $D_L$ and $D_T$ are  presented in Figs.~\ref{fig1}~(b)
and \ref{fig2}
as functions of $\beta$.
The mass $m$ for $T=0$ is in good agreement with the theoretical
prediction\cite{Polyakov} valid for a dilute monopole gas.
In the confinement phase, the anomalous dimensions $\alpha \ne 0$
for all propagators. They are functions not only of the monopole density
(which is monotonously decreasing with growing $\beta$). The cluster
structure of the monopole configurations seems to play a significant
r\^ole. The fit parameters $\alpha_L$ and $m_L$ of $D_L$ vanish at
$\beta_c$ and beyond,
a clear signal of the finite temperature phase transition\cite{CISLetter}
caused by dipole formation.
Due to multiple Dirac strings wrapping around the temporal direction
the behavior of $D_T$ in the deconfined phase (at larger $\beta$) is
different from $D_L$.
We carefully studied the Gribov copy dependence of the propagators.
The asymptotic
behaviors  of $D$ and $D_L$ are reached quickly,  contrary to the
$D_T$ propagator in the deconfinement. Further details can be found
in Ref.\cite{Chernodub:2002gp}

\end{document}